\documentclass[aps,final,twocolumn,showpacs,preprintnumbers,groupedaddress,superscriptaddress,amsmath,amssymb,amsfonts,floatfix]{revtex4}

\usepackage[]{graphicx}
\usepackage{dcolumn}
\usepackage{amsmath}
\usepackage{amssymb}
\usepackage{textcomp}
\usepackage{bm}

\begin{document}

\title{Melting of the orbital order in $ \textbf{LaMnO}_{\bf 3} $ probed by NMR}

\date{\today}

\author{A.~Trokiner}
\affiliation{LPEM, ESPCI ParisTech, UMR 8213, CNRS, 75005 Paris, France}

\author{S.~Verkhovskii}
\affiliation{LPEM, ESPCI ParisTech, UMR 8213, CNRS, 75005 Paris, France}
\affiliation{Institute of Metal Physics, Ural Branch of Russian Academy of Sciences, 620041 Ekaterinburg, Russia}

\author{A.~Gerashenko}
\affiliation{LPEM, ESPCI ParisTech, UMR 8213, CNRS, 75005 Paris, France}
\affiliation{Institute of Metal Physics, Ural Branch of Russian Academy of Sciences, 620041 Ekaterinburg, Russia}

\author{Z.~Volkova}
\affiliation{Institute of Metal Physics, Ural Branch of Russian Academy of Sciences, 620041 Ekaterinburg, Russia}

\author{O.~Anikeenok}
\affiliation{Institute of Physics, Kazan Federal University, 420008 Kazan, Russia}

\author{K.~Mikhalev}
\affiliation{Institute of Metal Physics, Ural Branch of Russian Academy of Sciences, 620041 Ekaterinburg, Russia}

\author{M.~Eremin}
\affiliation{Institute of Physics, Kazan Federal University, 420008 Kazan, Russia}

\author{L.~Pinsard-Gaudart}
\affiliation{Univ. Paris-Sud, Institut de Chimie Mol\'{e}culaire et des Mat\'{e}riaux d'Orsay, UMR8182, Bat. 410, Orsay 91405 France}
\affiliation{CNRS, Orsay, 91405 France}

\begin{abstract}

The Mn spin correlations were studied near the $O'$--$O$ phase transition at $T_{JT} =750{\rm~K}$, up to $ 950{\rm~K}$ with ${}^{17} {\rm O}$ and ${}^{139} {\rm La}$ NMR in a stoichiometric ${\rm LaMnO}_{{\rm 3}} $ crystalline sample. The measured local hyperfine fields originate from the electron density transfered from the $e_{g}^{} $- and $t_{2g}^{} $-orbitals to the 2\textit{s}(O) and 6\textit{s}(La) orbits, respectively. By probing the oxygen nuclei, we show that the correlations of the Mn spins are ferromagnetic in the {\textit{ab}}-plane and robust up to $T_{JT} $ whereas along the \textit{{c}}-axis, they are antiferromagnetic and start to melt below $T_{JT}$, at about  550~K. Above $T_{JT} $ the ferromagnetic Mn--Mn exchange interaction is found isotropic. The room temperature orbital mixing angle, $\varphi_{\rm nmr}= 109\pm 1.5{}^\circ $, of the $e_{g}^{} $ ground state is close to the reported value which was deduced from structural data on Jahn-Teller distorted $\rm MnO_6$ octahedra. For $T > T_{JT}$ $\rm LaMnO_3$ can be described in terms of non-polarized $e_g$-orbitals since both $e_g$-orbitals are equally occupied.

\end{abstract}


\maketitle

\section{Introduction}
The oxide ${\rm LaMnO}_{{\rm 3}} $ is a key system for experimental and theoretical studies that aim to resolve the relative importance of the electron-electron (\textit{e--e}) and electron-lattice (\textit{e--l}) interactions for the orbital physics of manganites \cite{Salamon_RevModPhys73}. The orbital degree of freedom originates from the singly occupied degenerate $e_{g}^{} $-state $\left(d_{3z^{2}-r^2 } \equiv {\left| \theta  \right\rangle} ,~d_{x^{2} -y^{2} } \equiv {\left| \varepsilon  \right\rangle} \right)$ of the Jahn-Teller (JT) active ${\rm Mn}^{{\rm 3}+} $ ($t_{2g}^{3} e_{g}^{1}$) ions \cite{Kugel_Khomskii}. Owing to the coupling of  orbital degrees of freedom with the lattice, at ambient conditions, in the $O'$-phase, the orthorhombic (\textit{Pbnm}) structure of ${\rm LaMnO}_{{\rm 3}} $ adopts a correlated pattern of corner-shared JT-distorted ${\rm MnO}_{{\rm 6}} $ octahedra with long (\textit{l}) and short (\textit{s}) $\rm Mn$--$\rm O$ bond lengths alternating in the {\textit{ab}}-plane. This structural signature of the long-range orbital order (OO) involves the low-lying orbital state ${\left| \psi _{g}  \right\rangle} =\cos (\varphi /2){\left| \theta  \right\rangle} +\sin (\varphi /2){\left| \varepsilon  \right\rangle} $ \cite{Kanamori_JAP31} which replicates the local symmetry of the oxygen environment at each Mn-site. Based on this correspondence, the room temperature value of the orbital mixing angle, $\varphi$, was estimated from $\rm MnO_6$ octahedron distortions,  $\varphi_{\rm str} \sim 108{}^\circ $ \cite{Rodr-Carv_PRB57}. This value is significantly smaller than $\ 120{}^\circ $, the prediction of the JT model \cite{Kanamori_JAP31}; this points out the important role of the superexchange (SE) mechanism \cite{Kugel_Khomskii} in the spatial ordering of the occupied $e_{g}^{} $-orbitals. The inclusion of the real structure of $\rm LaMnO_3$ into the dynamic mean-field theory (DMFT) calculations \cite{Leonov_PRB81,Pavarini_PRL104} has allowed getting $\varphi_{\rm dmft} \sim 109{}^\circ $ at room~$T$ \cite{Pavarini_PRL104} taking into account both, \textit{e--l} and \textit{e--e} interactions.

\begin{figure}[!b]
\includegraphics[width=0.75\hsize]{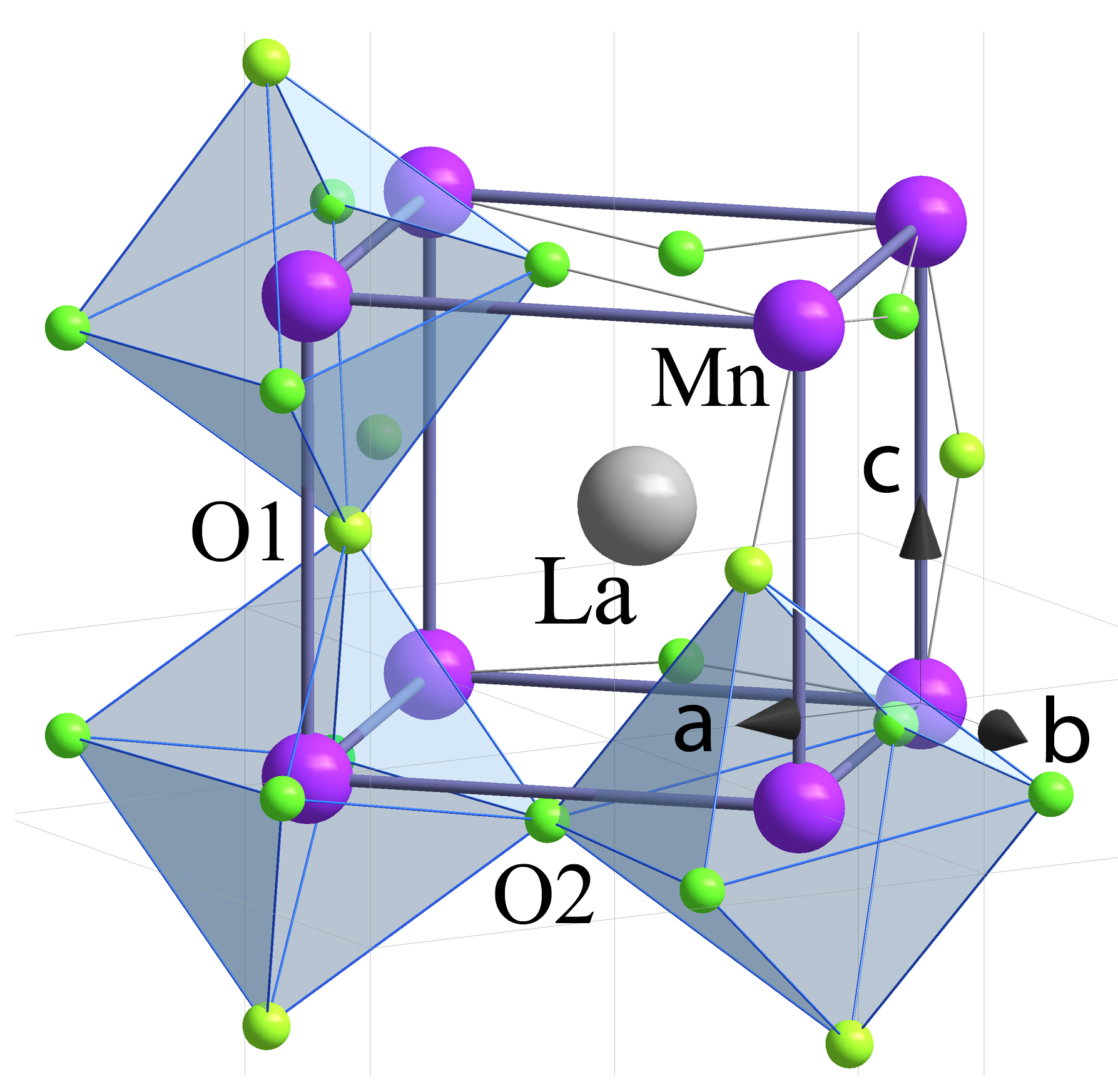}
\caption{\label{Fig. 1.}(Color online) Schematic structure (pseudo-cubic cell) of orthorhombic $\rm LaMnO_3$ showing O1 and O2 sites.}\end{figure}

According to DMFT, the $e_{g}^{} $ orbital state remains high-polarized even at $T>T_{JT} =750{\rm~K}$, where cooperative JT-distortions disappear \cite{Rodr-Carv_PRB57}. In contrast to this prediction, Raman spectroscopy evidences that orbital-disorder fluctuations are present well below $T_{JT}$ \cite{Granado_PRB62} indicating a thermal instability of OO in the $O'$-phase \cite{Martin-Carron}. The crystal structure of the high-temperature $O$-phase appears almost cubic on average \cite{Rodr-Carv_PRB57}, while dynamical JT distortions of the $\rm MnO_6$ octahedra remain up to about $1150$~K \cite{Sanchez_PRL90,Qiu_PRL94}. 

Unfortunately, to date experimental structural information on OO \cite{Rodr-Carv_PRB57,Granado_PRB62, Martin-Carron, Sanchez_PRL90,Qiu_PRL94} has not much electronic counterpart. The OO provides a remarkable anisotropy of the effective exchange interaction \cite{Kugel_Khomskii}, which explains the A-type antiferromagnetic (AF) spin order below $T_N\sim140~{\rm K}$ \cite{Goodenough_PR100,Moussa_PRB54}. Moreover, the estimates of the exchange integral values \cite{Moussa_PRB54} show that the AF exchange along the \textit{c}-axis is weaker than the ferromagnetic (FM) exchange in the \textit{ab}-layer. In the paramagnetic (PM) phase, this static spin order transforms into a time fluctuating short-range spin order of Mn neighbors so that no valuable information on anisotropy of the effective exchange interactions can be obtained above $T_N$ from bulk magnetic and transport measurements of intrinsically-twinned $\rm LaMnO_3$ crystals \cite{Zhou_Goodenough_PRB60}.

In this paper we resolve issues about the Mn--Mn spin correlations anisotropy and its variation across $O'$--$O$ transition in $\rm LaMnO_3$ by means of $\rm {}^{17}O$ NMR. In orthorhombic $\rm LaMnO_3$ there are two structural oxygen sites: O1 and O2 (Fig.~\ref{Fig. 1.}). The pathway of the SE interaction between two Mn neighbors involves O2 site in the \textit{ab}-plane and O1 along the \textit{c}-axis. The nuclear spin, ${}^{17}I$, probes the unpaired electronic spins on the $e_g$-orbitals through the transferred hyperfine interactions (THI) \cite{Yakubovskii_PRB74} which scalar part is almost independent of the Mn--$\rm Oi$--Mn bond bending. The scalar THI traces properties of the SE interaction at each site i.e. in the \textit{ab}-plane for O2 site and along the \textit{c}-axis for O1 site. The nuclear spin, ${}^{139}I$, of the La cation  probes the electron spin density of the $\rm Mn^{3+}$ ion transferred to the $6s({\rm La})$ orbital through the ${\rm Mn} (t_{2g})$--${\rm O2}(2p_{\pi})$--${\rm La}(6s)$ path \cite{Yoshinari_PRB60}. Because of its \textit{eight} $\rm Mn^{3+}$ neighbors the scalar THI traces the SE interaction averaged within the pseudo-cubic cell.

\section{Experimental}
The preparation and characterization of a stoichiometric $\rm LaMnO_3$ crystalline sample enriched by $\rm {}^{17}O$ NMR isotope up to about $5.5\%$, was described in \cite{L_Pinsard_Loreynne-Gaudar_SSC_151}. At room temperature, in the $Pbnm$ space group, the lattice parameters are: $a = 5.5379(1)$~\AA, $b = 5.7484(1)$~\AA, $c = 7.6950(1)$~\AA. 

The magnetic susceptibility $\chi = M/H$ was measured on a slice of the crystal from $T = 140$~K to 300~K with a SQUID magnetometer (Quantum Design) in a magnetic field $H = 50$~kOe and in the range $(295 - 810)$~K with a Faraday balance technique at $H = 4.5$~kOe.

The $\rm {}^{17}O$ and $\rm {}^{139}La$ NMR spectra (Fig.~\ref{Fig. 2.}) were acquired up to 950~K, in a crushed part ($\sim 200$~mesh) of the crystal, with a AVANCE III BRUKER spectrometer operating at $H = 11.7$~T. At this field, the Larmor frequency ($\nu_L$) is 67.800~MHz for $\rm {}^{17}O$ (standard $\rm H_2O$ liquid reference) and 70.647~MHz for $\rm {}^{139}La$.

As the quadrupolar interaction is present for $\rm {}^{17}O$ (${}^{17}I = 5/2$) and $\rm {}^{139}La$ nuclei (${}^{139}I = 7/2$), both spectra are very broad, approximately 2 and 25~MHz, respectively. Although the $\rm {}^{17}O$ and $\rm {}^{139}La$ spectra overlap, they can be separated and O1, O2 lines can be identified as was described in \cite{L_Pinsard_Loreynne-Gaudar_SSC_151}. As the spectra are broad, a method of frequency sweeping was used. A pulse sequence $\alpha-\tau-2\alpha-\tau-(echo)$  was used with a delay $\tau=12$~\textmu s and a pulse duration $\alpha\approx 1$~\textmu s, shorter than the one which optimizes the echo signal amplitude of both nuclei. The total spectrum was obtained by summing the Fourier-transformed half-echo signals acquired at equidistant  operating frequencies ($\rm step=0.1$~MHz). The simulation of the $\rm {}^{139}La$ quadrupolar split spectra, including both, the central ($m_I = -1/2 \leftrightarrow +1/2$) and the satellite transitions, was performed to determine at each temperature the components of the magnetic shift $\{K_{ii}\}$ as well as the quadrupole frequency $\nu_Q = 3eQV_{zz}/2I(2I-1)h$ and asymmetry parameter $\eta = |(V_{xx} - V_{yy})/V_{zz}|$ of the electric field gradient (EFG) tensor $\{V_{ii}\}$. The powder pattern simulation program takes into account the quadrupole coupling corrections up to the second order in $\sim\nu_Q/\nu_L$.

The $\rm {}^{17}O$ spin echo decay rate, ${}^{17}T_2^{-1}$, was measured on the peak of each $\rm {}^{17}O$ NMR line. The echo-decay data were collected by varying $\tau$. The characteristic time of the echo-decay, ${}^{17}T_2$, is defined as the time at which the echo-signal $E(2\tau)$ drops to $1/e$ of its starting value.

\section{Results and discussion}
\subsection{Spin and charge environment of La: magnetic susceptibility and $\rm \bf {}^{139}La$ NMR}
\begin{figure}[!t]
\includegraphics[width=0.70\hsize]{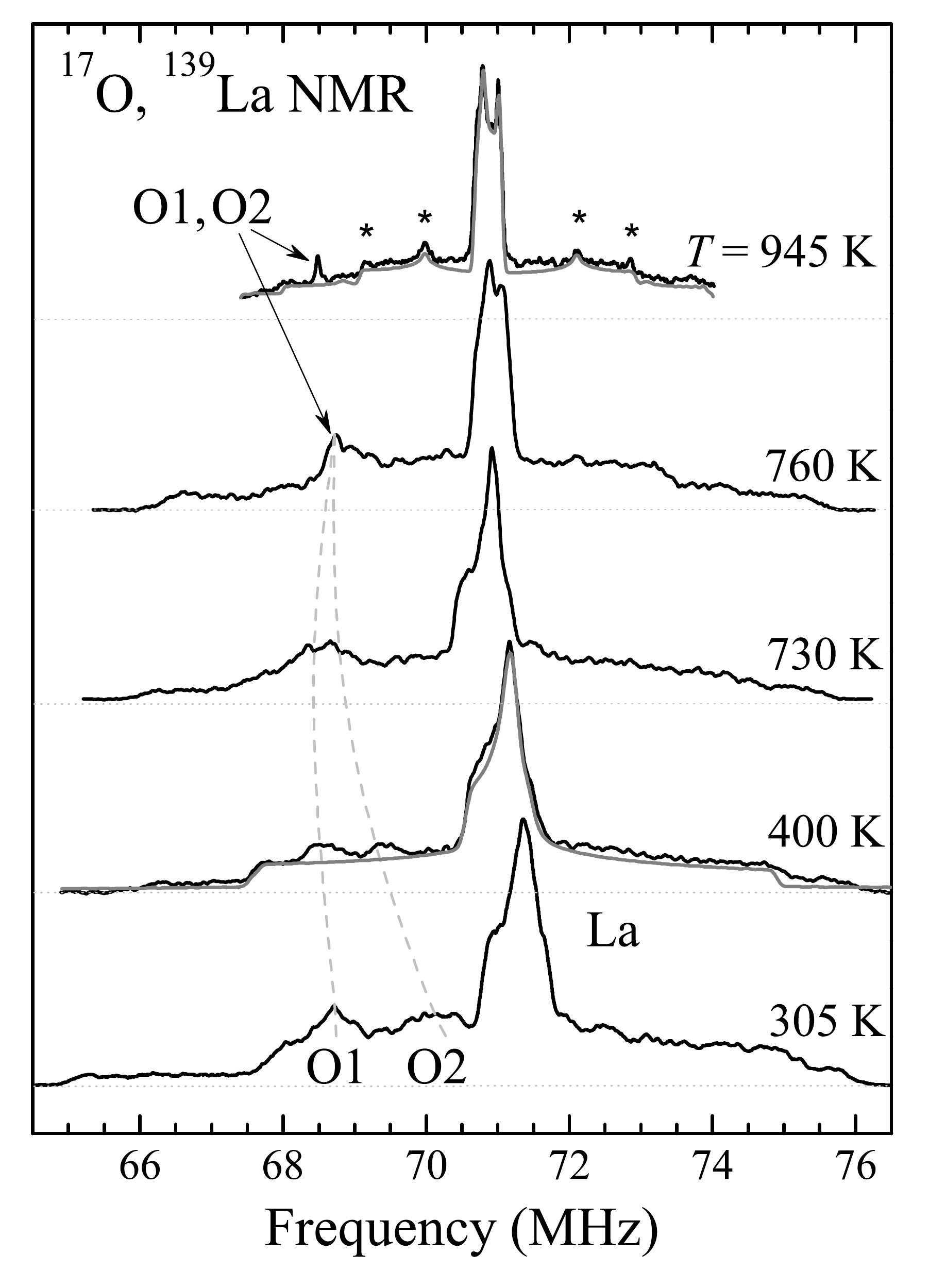}
\caption{\label{Fig. 2.}The ${}^{17} {\rm O}$ and ${}^{139} {\rm La}$ NMR powder spectra of stoichiometric ${\rm LaMnO}_{3} $ ($T_{JT} =750{\rm~K}$). The low frequency $\rm {}^{139}La$ satellite lines overlap with the O1 and O2 lines.  The dotted lines joining O1 peaks and O2 peaks are guide for the eyes. At 945~K, the narrowed satellite peaks of the $\rm {}^{139}La$ NMR spectrum are marked by ($\ast$). The grey curve beneath the spectrum at $T = 400$~K and 945~K is the corresponding $\rm {}^{139}La$  NMR spectrum simulation.}\end{figure}

\begin{figure}[!b]
\includegraphics[width=0.85\hsize]{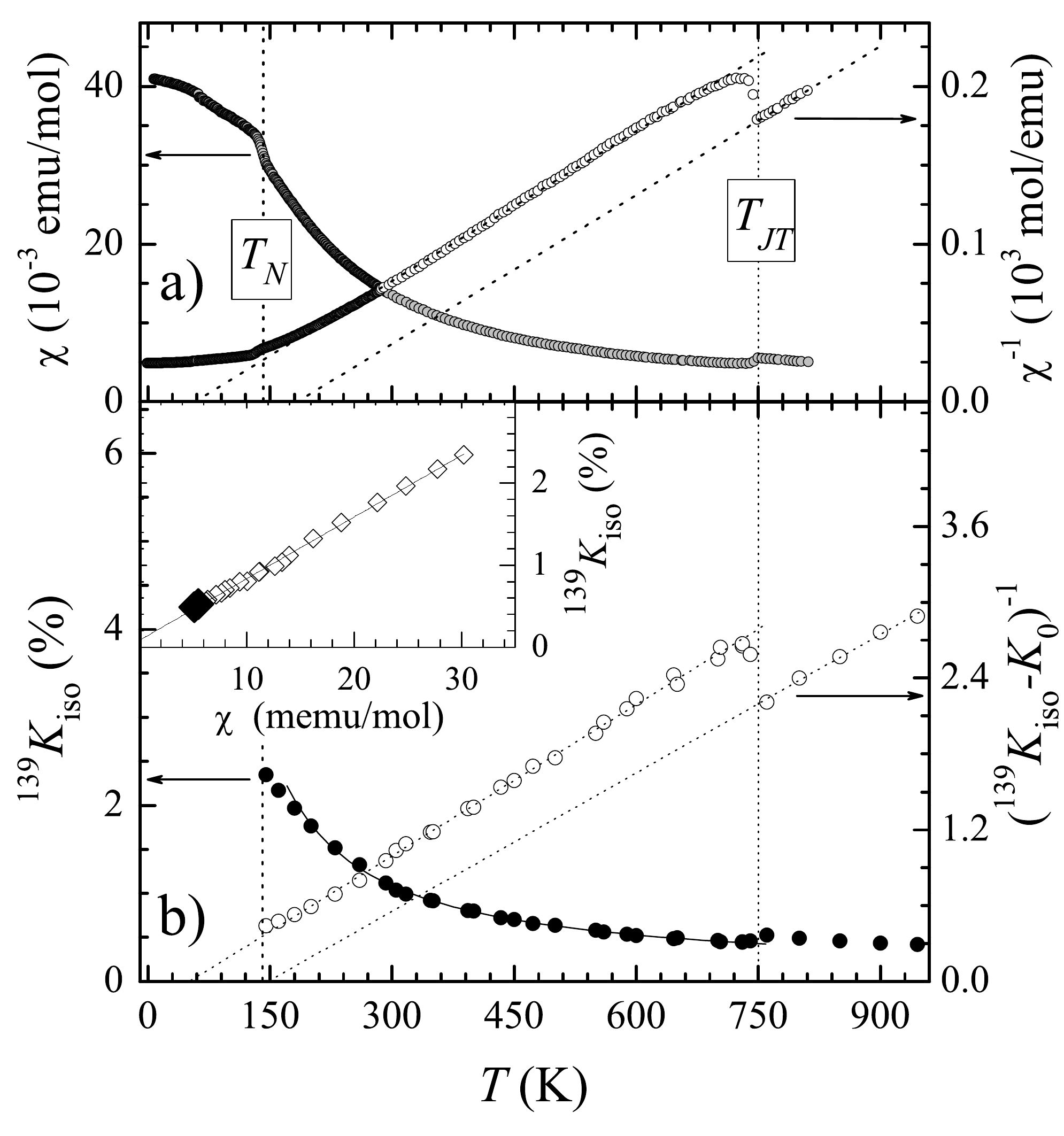}
\caption{\label{Fig. 3.}(a) Magnetic susceptibility $\chi =M/H$ vs \textit{T} of ${\rm LaMnO}_{3} $ crystal. The doted lines are the fits to $\chi ^{-1} (T)$. (b) $\rm {}^{139}La$ isotropic shift, $^{{\rm 139}} K_{{\rm iso}} $  vs \textit{T} ($\bullet$) and inverse of (${}^{139}K_{\rm iso} - K_0$) vs $T$ ($\circ$). The solid curve is the Curie-Weiss fit for $T<750{\rm~K}$, the dotted lines are the fits to $({}^{139}K_{\rm iso} - K_0)^{-1}$. Inset: $^{{\rm 139}} K_{{\rm iso}} (\chi )$ plot below 750~K ($\lozenge$) and above 750~K ($\blacklozenge$).}\end{figure}

Figure~\ref{Fig. 3.}a shows the thermal variation of the magnetic susceptibility of a slice of the crystal. The two phase transitions are clearly seen: the onset of the AF ordered phase is displayed as a change of the $\chi (T)$ slope at $T_N \sim 139$~K whereas the upturn at $T_{JT} = 750(2)$~K indicates the JT transition.  In the $O'$-phase, from 250~K up to 650~K, the magnetic susceptibility follows a Curie-Weiss (C-W) law $\chi(T) = \chi_0 + C/(T - \Theta)$ with a Weiss temperature $\Theta(O'\text{-phase}) = 67(5)$~K and negligible value of $\chi _{0} \approx 0.1~{\rm  memu/mol}$. The Curie constant $C=3.17(10)~{\rm  emu}\cdot {\rm K/mol}$ corresponds to an effective magnetic moment $\mu_{\rm eff} = 5.04~\mu_B$ which value is only slightly larger than the expected value $\mu_{\rm eff} = 2 \mu_B (S (S + 1))^{1/2} =  4.89~\mu_B$ for the ${\rm Mn^{3+}_{}}(t^3_{2g}e^1_g,{}^{5}E)$ state. Above $T_{JT}$ the C-W law fits well $\chi(T)$ with almost the same $C$ and $\chi_0$ value but with a large positive $\Theta(O\text{-phase}) = 197(5)$~K. The significant increase of $\Theta$ on crossing $T_{JT}$ indicates that the Mn--Mn spin correlations are FM-enhanced in the $O$-phase in agreement with previous $\chi$ results \cite{Zhou_Goodenough_PRB60}. 

The magnetic shift and EFG parameters of $\rm {}^{139}La$ nucleus were deduced from the simulation of the $\rm {}^{139}La$ NMR spectra which are represented on Fig.~\ref{Fig. 2.} for two selected temperatures, 400 and 945~K. Compared to the $O'$-phase, the double peaked $\rm {}^{139}La$ central line in the $O$-phase reflects the decrease of $\eta$, the asymmetry parameter.

The isotropic magnetic shift, ${}^{139}K_{\rm iso} \equiv 1/3Tr\{K_{ii}\}$, scales well the thermal behavior of $\chi$ in both phases (Fig.~\ref{Fig. 3.}b). Indeed, the corresponding local field, $h_{\rm loc}({\rm La})= {}^{139}K_{\rm iso}H$, is caused by the Fermi-contact interaction of ${}^{139}I$ with the electron spin density $f_{s,{\rm La}}\langle s_z({\rm Mn})\rangle$ transferred at the $6s({\rm La})$ orbital from the eight $\rm Mn^{3+}$ neighbors mainly through the ${\rm Mn}(t_{2g})$--${\rm O2}(2p_\pi)$--${\rm La}(6s)$ path so that $h_{\rm loc}({\rm La}) = 8 f_{s,{\rm La}} H_{FC}(6s) \langle s_z({\rm Mn})\rangle$. Compared with the bulk susceptibility $\chi$, ${}^{139}K_{\rm iso}$ is proportional to the thermal averaged projection of the Mn spin $\langle s_z({\rm Mn})\rangle \sim (T-\Theta_{\rm La})^{-1}$, which reflects the net $t_{2g}$ spin polarization of Mn within each pseudo-cubic unit cell. ${}^{139}K_{\rm iso}(T)$ follows also a C-W law: $K_0 +C_{\rm La}/(T-\Theta_{\rm La})$ with a $T$-independent term, $K_0 = 0.07(4)\%\ll {}^{139}K_{\rm iso}(T)$ and $\Theta_{\rm La} (O')= 55(12)$~K and $\Theta_{\rm La} (O)= 160(40)$~K. The $\rm {}^{139}La$ NMR data confirm that the Mn--Mn exchange coupling becomes more ferromagnetic in the $O$-phase. 

\begin{figure}[!t]
\includegraphics[width=0.87\hsize]{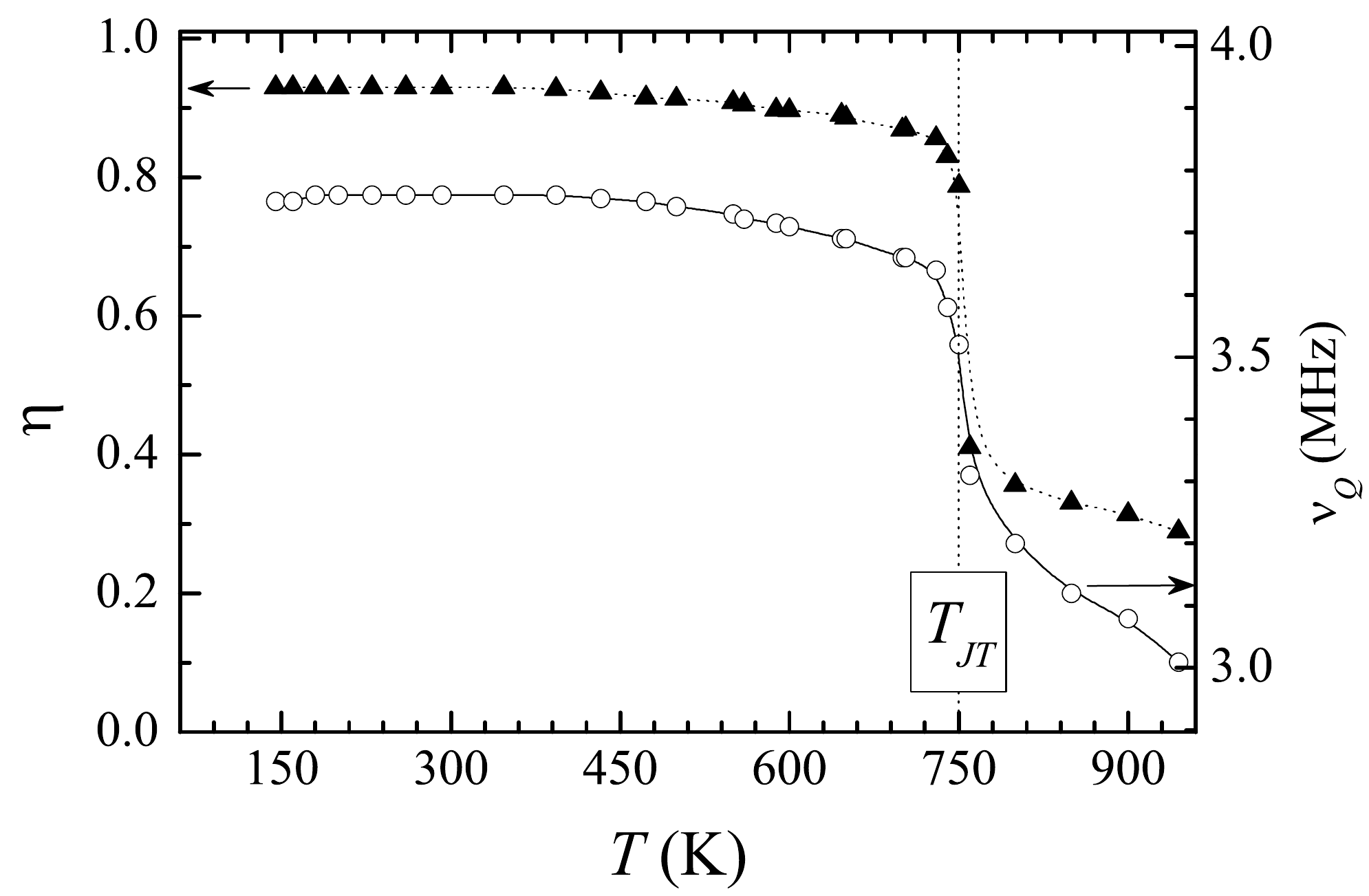}
\caption{\label{Fig. 4.}Thermal behavior of the quadrupolar frequency, ${}^{139}\nu_Q$ ($\circ$), and EFG asymmetry parameter, ${}^{139}\eta$ ($\blacktriangle$), at La site.}\end{figure}

Fig.~\ref{Fig. 4.} concerns the charge environment of La cation. In the $O'$-phase $\nu_Q$ and $\eta$ remain practically unchanged up to about 500~K, smoothly decrease at higher temperature and drop on crossing $T_{JT}$. The non zero value of the asymmetry parameter, $\eta \approx 0.3$, is in agreement with the symmetry of the $O$-phase which was found still orthorhombic in the $O$-phase \cite{Rodr-Carv_PRB57}. The drop of $\nu_Q$ and $\eta$ on crossing $T_{JT}$, is due to the fact that although there is no change in symmetry, the $\rm MnO_6$ octahedra become more regular in the $O$-phase and the observed lattice appears cubic ($a \approx b \approx c /\sqrt{2}$) \cite{Rodr-Carv_PRB57}.

It is worth to mention that close to $T_{JT}$, in the $O'$ and $O$-phases, all La sites are identical since a single set of the $\rm {}^{139}La$ NMR parameters, $\{K_{ii}; V_{ii}\}$, is enough to describe the magnetic and charge environment of La. Therefore, static nanoscale heterogeneities which might appear in $\rm LaMnO_3$ \cite{Zhou_Goodenough_2003} when approaching $T_{JT}$ from below are not confirmed by $\rm {}^{139}La$ NMR.

\subsection{Anisotropy of the Mn--Mn exchange interaction: $\rm \bf {}^{17}O~NMR$}
${}^{17} {\rm O}$ NMR shed light on the Mn--Mn pair spin correlations which are mainly related to $e_{g}^{} $-electron. The covalent electron transfer from its two Mn nearest neighbors creates a fraction of spins, $f_s$, on the O($2s$) orbital. The hyperfine interaction of ${}^{17} I $ with the spin density $f_{s} \left\langle s_{{\rm z}} ({\rm Mn})\right\rangle $ results in a local field, ${}^{17} h_{{\rm loc}} $, which is responsible for the large positive shift of the ${}^{17} {\rm O}$ NMR lines. Below 750~K, the NMR powder spectrum consists of two lines, O1, O2 (Fig.~\ref{Fig. 2.}). The peak position, $\nu _{{\rm p}}$, of the ${}^{17} {\rm O}$ lines defines their isotropic shift ${}^{17} K= {}^{17}h_{\rm loc}/H=(\nu _{{\rm p}} -{}^{17} \nu _{L} )/{}^{17} \nu _{L} $, where ${}^{17}\nu _{L}$ is the Larmor frequency.

\begin{figure}[!t]
\includegraphics[width=0.85\hsize]{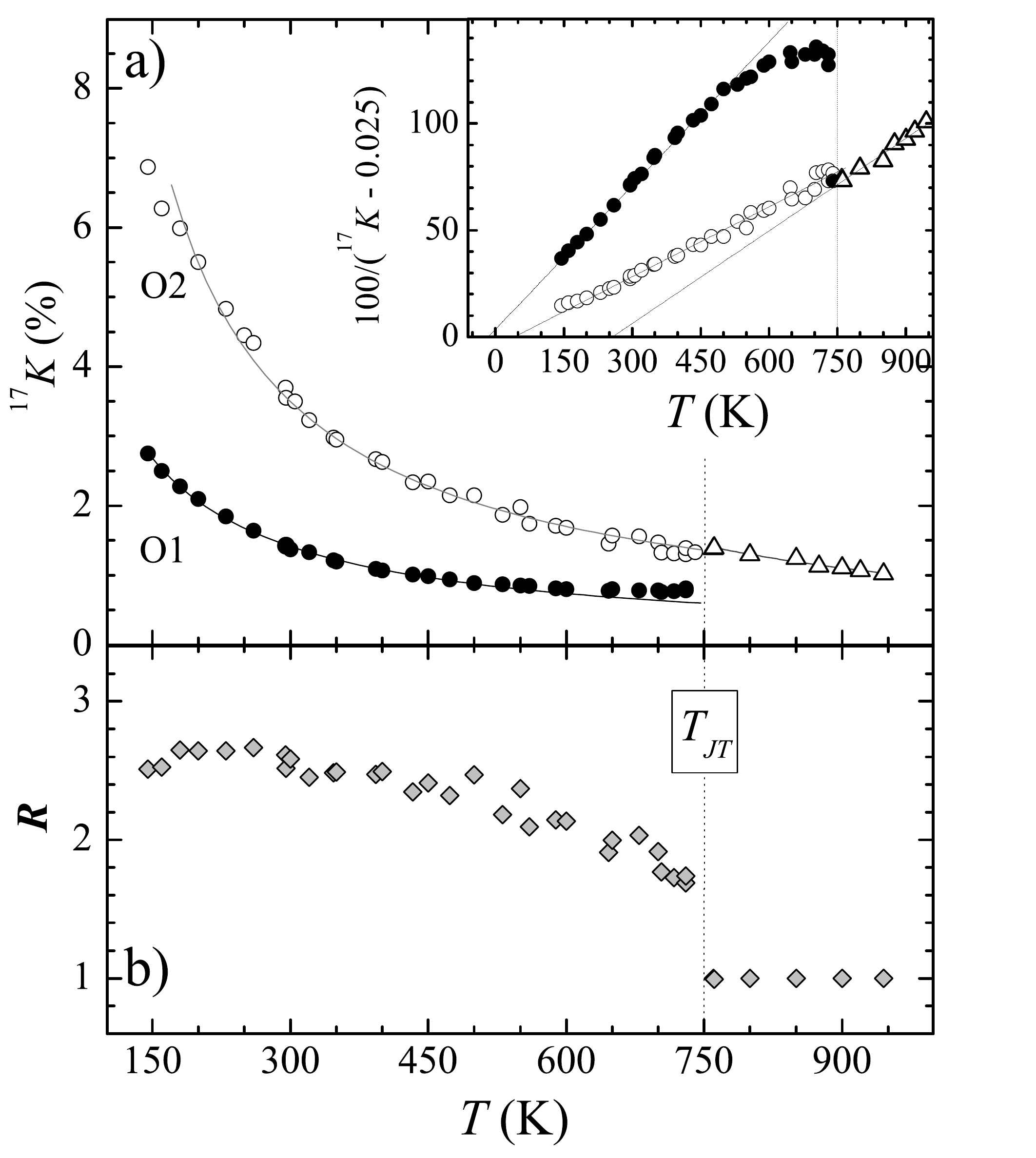}
\caption{\label{Fig. 5.}(a)Oxygen isotropic magnetic shift, ${}^{17}K$ vs $T$ for O1 ($\bullet$) and O2 ($\circ$) sites in the $O'$-phase and for the single magnetic $O$-site ($\vartriangle$) in the O-phase. Inset: inverse shift $\left\lbrace {}^{17} K-{}^{17} K_{0} \right\rbrace ^{-1} $ vs $T$. The solid curves are the Curie-Weiss fit to O1 $\left(180~{\rm K} < T < 550~{\rm K}\right)$, O2 $\left(180~{\rm K}< T < 750~{\rm K}\right)$ and to the single magnetic $O$-site above 750~K. (b) Thermal behavior of $R$, the ratio of the local field created at O2 and at O1 sites by the two Mn nearest neighbors (see text).}\end{figure}

\begin{figure}[!t]
\includegraphics[width=0.97\hsize]{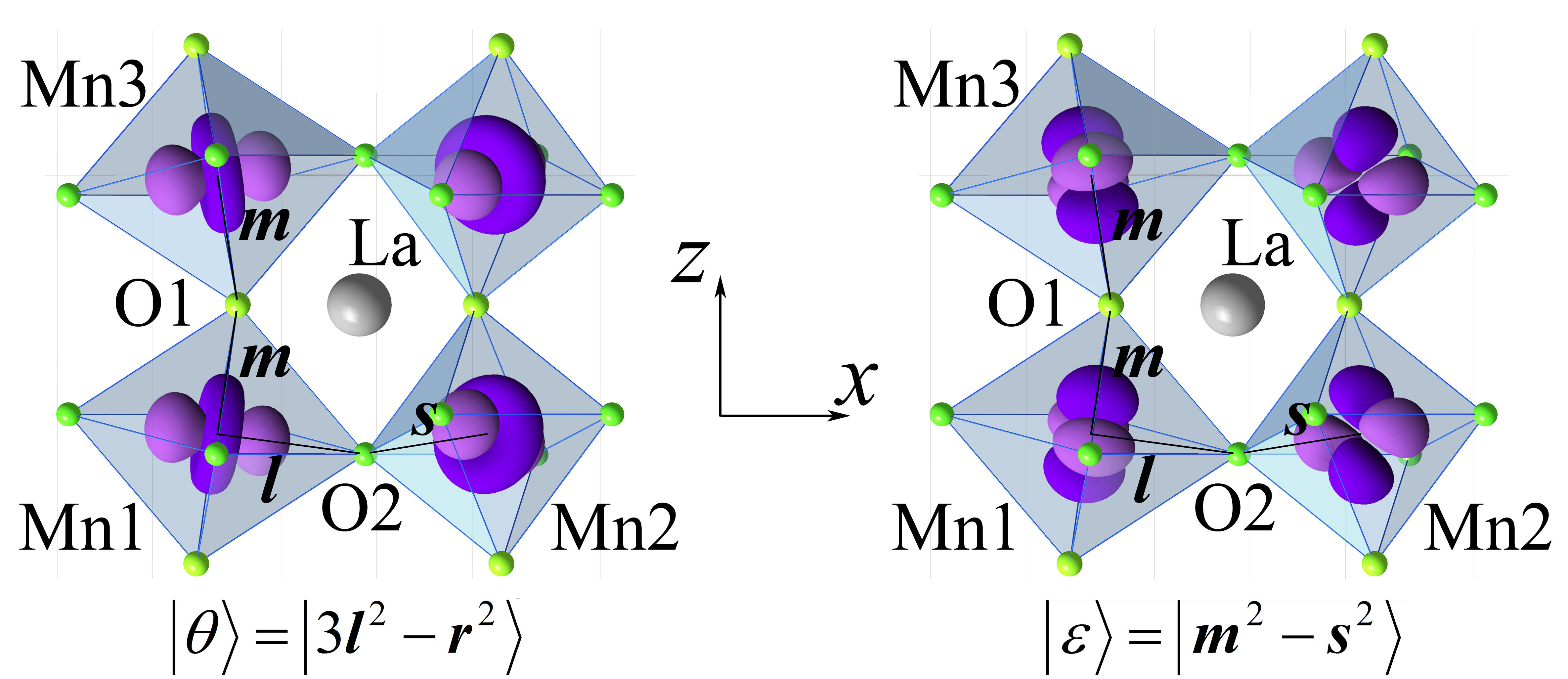}
\caption{\label{Fig. 6.}(Color online) Direction of the $e_g$-orbitals, ${\left| \theta  \right\rangle} $ and ${\left| \varepsilon  \right\rangle} $ in the JT distorted $\rm MnO_6$ octahedra. \textit{l}, \textit{m} and \textit{s} are the  long, medium and short  Mn--O bond lengths, respectively, $x$ and $z$ are pseudocubic axes.}\end{figure}

The line displaying the largest $^{17} K$ was attributed to oxygen atoms in O2 site   \cite{L_Pinsard_Loreynne-Gaudar_SSC_151}. From 200~K up to $T_{JT}$, ${}^{17}K(T;{\rm O2})$ data   follow a C-W law: $K_{0} +C_{\rm O2}/\left(T-\Theta _{{\rm O2}}\right) $ with a chemical shift $K_{0} \left({\rm O}2\right)=0.025\left(10\right){\rm\% }$ and $\Theta _{{\rm O2}} =35(10){\rm~K}$ (Fig.~\ref{Fig. 5.}a). The positive value of $\Theta _{{\rm O2}}$  evidences the FM nature of SE interaction between Mn neighbors in the {\textit{ab}}-plane. In contrast to ${}^{139} \rm {La}$ and ${}^{17} {\rm O2}$, the mean field law fits ${}^{17} K(T;{\rm O}1)$ data only up to  $T^{\ast}\sim 550$~K (inset Fig.~\ref{Fig. 5.}a), with $\Theta _{{\rm O1}} =-15(20){\rm~K}$ and $K_{0} ({\rm O1})=K_{0} ({\rm O}2)$. The slightly negative value of $\Theta _{{\rm O1}}$ demonstrates the AF nature of the interaction between adjacent Mn along \textit{c} (Mn1 and Mn3 in Fig.~\ref{Fig. 6.}) resulting from a small imbalance between several SE interaction involving $t_{2g}$ and $e_g(\theta, \varepsilon)$ orbitals \cite{Zhou_Goodenough_2006}. Above $T^{\ast}$, ${}^{17}K (T;{\rm O1})$ deviates from the C-W law  and tends towards ${}^{17}K (T;{\rm O2})$. The merging of O1 and O2 NMR lines at the $O'$--$O$ transition indicates that the two structurally distinct O sites have the same magnetic environment. Above 750~K, the C-W fit of $\{{}^{17} K(T)-{}^{17} K_{0} \} ^{-1} $ yields $K_{0} =0.025\left(10\right){\rm\% }$ and $\Theta _{{\rm O1, O2}} =260(50){\rm~K}$. Thus $\rm {}^{17}O$ NMR demonstrates that the Mn--Mn exchange interaction changes from anisotropic $\left(\Theta_{\rm O2}= 35~{\rm K},~\Theta_{\rm O1}= - 15~{\rm K}\right)$ to isotropic $\left(\Theta_{\rm O2}= \Theta_{\rm O1} = 260~{\rm K}\right)$ at $T_{JT}$ and that in the $O$-phase, the isotropic exchange coupling  is strongly FM enhanced.

Figure~\ref{Fig. 5.}b shows  the thermal variation of $R={}^{17} h_{{\rm loc}} ({\rm O2})/{}^{17} h_{{\rm loc}} ({\rm O1})$, the ratio of the local field at O2 and O1 sites. $R$ is mainly controlled by the orbital mixing angle $\varphi$. In a picture of localized $d$-electrons, the ground-state wave function of the $e_g$-electron can be represented as a linear combination of the $e_g$ atomic orbitals: ${\left| \psi_g  \right\rangle} =\cos (\varphi /2){\left| \theta  \right\rangle} +\sin (\varphi /2){\left| \varepsilon  \right\rangle} $ for each $\rm Mn^{3+}$ site \cite{Kugel_Khomskii} with $\varphi$ defined in the pseudocubic axes \textit{{x}}, \textit{{y}}, \textit{{z}} (Fig.~\ref{Fig. 6.}). We use also the wave function defined in the local coordinates of each Mn site, ${\left| \eta  \right\rangle} =c_1^{}{\left| \theta  \right\rangle} +c_2^{}{\left| \varepsilon  \right\rangle} $, e.g.  the ${\left| \theta  \right\rangle} $ orbital points in the \textit{{x}} direction for Mn1 site. In order to deduce $\varphi$, we have calculated ${}^{17}h_{\rm loc}({\rm Oi})$ by using an~effective Hamiltonian of the super-transferred electron-nuclear interaction:
\begin{eqnarray} \label{GrindEQ__1_}
H_{{\rm eff}} =\frac{16\pi {}^{17} \gamma \hbar \mu_{B} }{3} \sum _{\eta =\theta ;\varepsilon }a_{\eta }^{+} a_{\eta }^{} \left({\left\langle \eta  \mathrel{\left| \vphantom{\eta  2s}\right.\kern-\nulldelimiterspace} 2s \right\rangle} +\gamma _{\eta 2s} \right) \times &
\nonumber \\
{\left\langle 2s \right|} \textbf{{IS}}{\left| 2s \right\rangle}  \left({\left\langle 2s \mathrel{\left| \vphantom{2s \eta }\right.\kern-\nulldelimiterspace} \eta  \right\rangle} +\gamma _{\eta 2s} \right),
\end{eqnarray}
where the covalent transfer of unpaired electron density is accounted by  ${\left\langle 2s  | \eta \right\rangle} $, the overlap integrals, and by $\gamma _{\eta 2s } $, the covalence parameter. All overlap integrals (Table~I) were calculated by using Hartree-Fock's wave functions \cite{Clementi_Roetti} and atomic positions \cite{Rodr-Carv_PRB57} of the ${\rm Mn}^{{\rm 3}+} $ and ${\rm O}^{2-} $ ions. As illustrated in Fig.~\ref{Fig. 6.},  O1 atom is coupled to Mn1, Mn3 ions via the middle ($m$) bonds, whereas O2 is connected to Mn1, Mn2 neighbors in {\textit{ab}}-plane via the long ($l$) and short ($s$)  Mn--O bonds. The covalence parameters $\gamma _{\eta 2s} $ were assumed to be proportional to the corresponding overlap integrals: $A_{\alpha}\equiv {\left\langle 2s \mathrel{\left| \vphantom{2s \theta }\right.\kern-\nulldelimiterspace} \theta  \right\rangle_{\alpha}}$, $B_{\alpha}\equiv {\left\langle 2s \mathrel{\left| \vphantom{2s \varepsilon }\right.\kern-\nulldelimiterspace} \varepsilon  \right\rangle_{\alpha}}$ $\left(\alpha=l,~m,~s\right)$.

Finally, the \textit{R}-ratio takes the form: 
\begin{equation} \label{GrindEQ__2_} R =\frac{c_{1}^{2} A_{l}^{2} +c_{1}^{2} A_{s}^{2} + c_{2}^{2} B_{s}^{2} +2 A_{s} B_{s} c_{1} c_{2} }{2\left[c_{1}^{2} A_{m}^{2} + c_{2}^{2} B_{m}^{2} +2 A_{m} B_{m} c_{1} c_{2} \right]},  
\end{equation} 
where the terms $ c_{{\rm 1}}^{2}{ A_{\alpha}^2},~c_{{\rm 2}}^{2}{ B_{\alpha}^2} $ are proportional to the conventional spin densities $f_{s,\theta},~f_{s,\varepsilon} $ \cite{Owen_RPP29} transferred from $\rvert \theta \rangle $ and $\rvert \varepsilon \rangle$ to O($2s$) orbital and the cross term, $\propto c_{1}^{} c_{2}^{} $, describes the quantum interference effect of the interaction~\eqref{GrindEQ__1_}. The orbital mixing coefficients $c_{1}^{} $, $c_{2}^{} $ and  $\varphi $ were estimated with eq.~\eqref{GrindEQ__2_} and their value for Mn1 site are listed at 293~K, 573~K and 798~K in Table~I. At 293~K, $R=2.5\pm 0.1$ yields $\varphi _{\rm{nmr}} =109 \pm1.5^\circ$, close to $\varphi_{\rm str} \sim 108^\circ$ deduced from structural data on Jahn-Teller distorted $\rm MnO_6$ octahedra in $\rm LaMnO_3$ \cite{Rodr-Carv_PRB57}. Besides, the cross term $c_{1}^{} c_{2}^{} $ is negative, in agreement with the orthorhombic crystal-field parameters ratio, $E/D={c_{1}^{} c_{2}^{}/ \left(c_{2}^{2} -c_{1}^{2} \right)} $, which was found positive in ${\rm La} {\rm MnO}_{{\rm 3}} $ \cite{Matsumoto} and in an untwined ${\rm La}_{{\rm 0.95}} {\rm Sr}_{{\rm 0.05}} {\rm MnO}_{{\rm 3}} $ ($T_{JT} {\rm \approx 605\; K}$) single crystal  \cite{Deisinhofer_PRB65}. For $T>T_{JT} $, the fast  fluctuations ($t_{s} \sim 10^{-14} {\rm~sec}$) of the JT-distorted octahedra \cite{Granado_PRB62,Martin-Carron} are averaged over the time scale of NMR spectra ($t_{{\rm nmr}} >10^{-7} {\rm~sec}$). Thus, we measure the time-average local magnetic field, $\left\langle {}^{17} h_{{\rm loc}} ({\rm Oi})\right\rangle _{\displaystyle{{t}_{\rm{ nmr}}}}$. Due to the $\rm MnO_6$ octahedra fast fluctuations during $t_{{\rm nmr}}$, the coherent cross term $ \left\langle c_1^{}c_2^{}\right\rangle _{\displaystyle{{t}_{\rm{ nmr}}}}$ vanishes in eq.~\eqref{GrindEQ__2_} yielding $c_1^{2}\approx c_2^{2}=0.50(2)$ for $R(798~{\rm K}) =1.0$. Therefore, for $T>T_{JT}$ and at the time scale $t_{\rm{ nmr}}$, $\rm LaMnO_3$ can be described in terms of non-polarized $e_g$-orbitals, since $|\theta\rangle $ and $|\varepsilon \rangle$ orbitals are equally occupied, in agreement with Ref. \cite{Rodr-Carv_PRB57}.

\begin{table}[t]
\caption{\label{Table I}Overlap integrals ($\times 10^{-2}$) between Mn(3\textit{d}) and O(2\textit{s}) orbitals  calculated for Mn1 site. Indexes $l$(\textit{{x}}), $m$(\textit{{z}}) and $s$(\textit{{y}}) refer to long, middle and short Mn1--O bond, respectively. $\varphi_{\rm nmr}=2\pi /3+2\arctan (c_2^{}/c_1^{})$.}
\begin{ruledtabular}
\begin{tabular}{ccccccccc}
\textit{T}(K)&$A_{l}$&$A_{m} $&$A_{s} $&$B_{m} $&$B_{s} $&$c_{1}^{} $&$c_{2}^{} $& \begin{tabular}{c}$\varphi _{{\rm nmr}} $\\(degree)\\\end{tabular} \\
\hline\\
293 & 3.843 & -2.904 & -3.263 & -5.031 & 5.651  & 0.995 & -0.10 & 109(3) \\
573 & 3.967 & -2.855 & -3.208 & -4.945 & 5.557  & 0.998 & -0.06 & 114(3) \\
798 & 5.117 & -2.682 & -2.795 & -4.645 & 4.840  & \multicolumn{2}{c} {$c^{}_{1}\approx c^{}_{2}$}  & $\sim90$ \\[\smallskipamount]
\end{tabular}
\end{ruledtabular}
\end{table}

\subsection{Low-frequency dynamics of Mn spin near $\bf {\textit T}_{\textit {JT}}$}

The low-frequency dynamics of the Mn spins was studied by measuring the ${}^{17} {\rm O}$ spin echo decay rate, $T_{2}^{-1} $, on  O1 and O2 lines.  $T_{2}^{-1} $ probes the time-dependent fluctuations of the local field ${}^{17}h_{\rm loc}$ at $\rm {}^{17}O$ nuclei. The transverse and longitudinal components of ${}^{17}h_{\rm loc}$, $h_\perp$ and $h_{\parallel}$, are defined as a function of the direction of $H$. In general both $h_\perp$ and $h_{\parallel}$ contribute to the echo-decay process \cite{Ch_Slichter,Walstedt_PRL19}:
\begin{equation} \label{GrindEQ__3_} 
T_2^{-1}(T) = {}^{17}\gamma^2 h^2_{\parallel}\tau_c + \left({}^{17}I +\frac{{}_1}{{}^2}\right)^2 T_1^{-1}(T),  
\end{equation}
where the nuclear spin-lattice relaxation rate, $T_1^{-1}$, involves only the transverse components, probing $\langle h_\perp(0)h_{\perp}(t)\rangle$. Usually at elevated temperature far above the magnetic transition, the second term in \eqref{GrindEQ__3_} dominates and the  contribution to the $\rm {}^{17}O$ echo decay rate \cite{Moriya_PrThPh16} at Oi site is:
\begin{eqnarray} \label{GrindEQ__4_} 
T_{2,{\rm i}}^{-1}(T)\propto \left({}^{17}I +\frac{{}_1}{{}^2}\right)^2  
\left[ f_sH_{FC}(2s)A_{\rm q}({\rm Oi})\right]^2 T\chi_{\rm i} (T),  
\end{eqnarray}
where $\chi_{\rm i}$ is the local spin susceptibility of Mn ions in the {\textit{ab}}-plane ($\rm i=2$) and along \textit{{c}}-axis ($\rm i=1$), it is proportional to ${}^{17}K(T;{\rm Oi})$. The factor $f_sH_{FC}(2s)$ is the local field created at the $\rm {}^{17}O$ nucleus by the unpaired spin of one Mn neighbor, $f_s$ is the spin density transferred from $e_g$ to ${\rm O}(2s)$ orbital and $H_{FC}(2s)$ is the Fermi contact field of the ${\rm O}(2s)$ orbital. The form-factors $A_{\rm q}({\rm Oi})$ takes into account the FM correlations of the Mn neighbours in the {\textit{ab}}-plane ($A_{\rm q}^2({\rm O2}) = 1$) and the AF correlations along the \textit{{c}}-axis ($A_{\rm q}^2({\rm O1}) < 1$). 

\begin{figure}[!t]
\includegraphics[width=0.9\hsize]{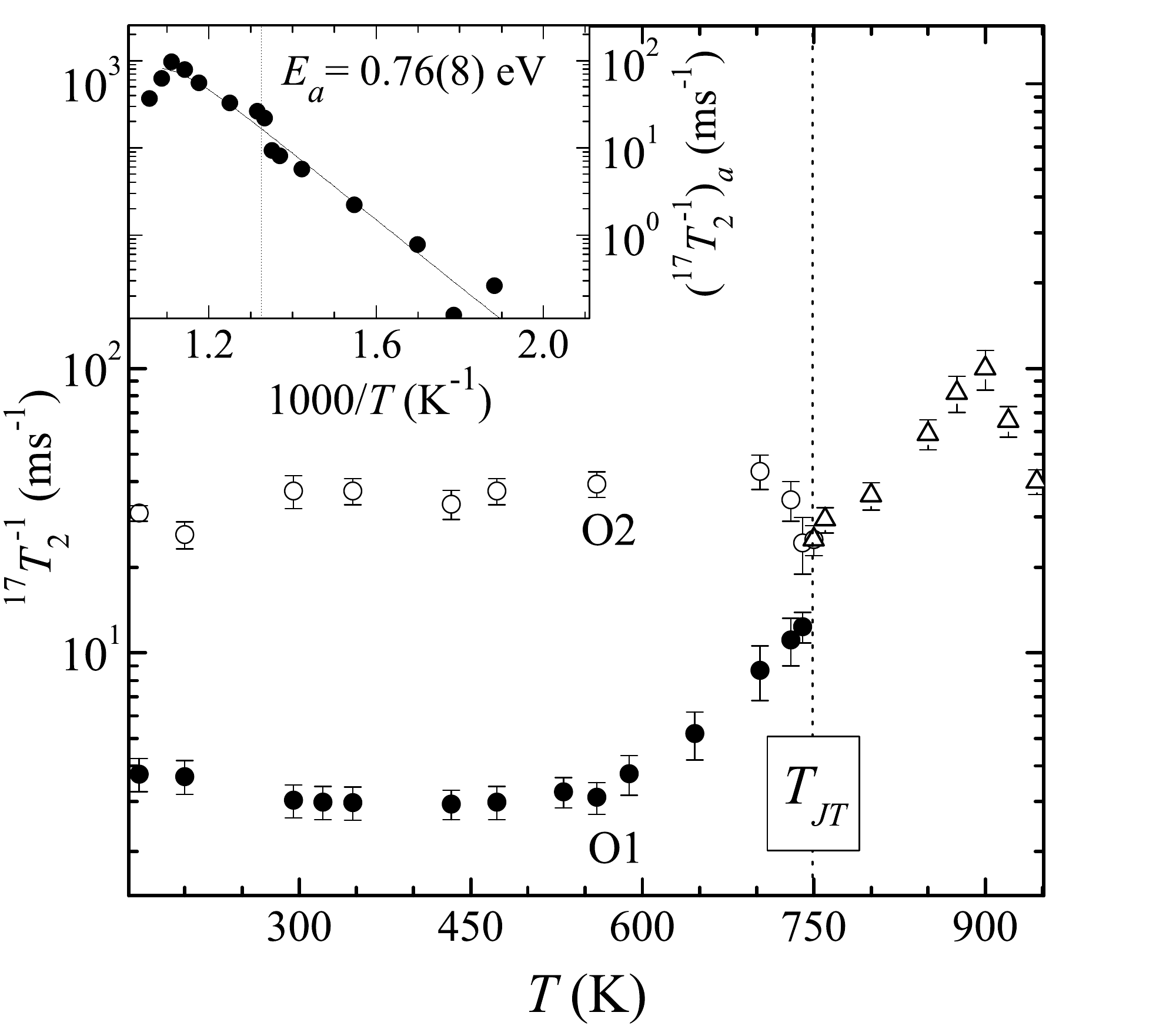}
\caption{\label{Fig. 7.}Thermal behavior of $\rm {}^{17}O$ echo decay rate, ${}^{17}T_2^{-1}$ for O1 ($\bullet$) and O2 ($\circ$) sites in the $O'$-phase and for the single magnetic $O$-site ($\vartriangle$) in the $O$-phase. Inset: thermally activated component $({}^{17}T_2^{ -1})_a^{} = \{{}^{17}T_2^{-1}(T) - {}^{17}T_2^{-1}(293~{\rm K})\}$ versus $1000/T$.}\end{figure}

The data on Figure~\ref{Fig. 7.} show that as long as the mean field behavior holds for ${}^{17}K(T;{\rm Oi})$ i.e. as long as $\chi_{\rm i} (T)$ is proportional to $(T- \Theta_{\rm i})^{-1}$, $T_2^{-1}({\rm Oi})$ is almost $T$-independent above room temperature in agreement with expression~(\ref{GrindEQ__4_}). Indeed, $T_2^{-1}({\rm O2})$ is constant up to $T\approx T_{JT}$ whereas $T_2^{-1}({\rm O1})$ is constant only up to 560~K that is, up to $\sim T^{\ast}$. Above $T^{\ast}$, $T_2^{-1}({\rm O1})$ starts to increase and approaches $T_2^{-1}({\rm O2})$ values close to $T_{JT}$. In the $O$-phase, $T_2^{-1}$ displays a maximum near 900~K. This behaviour reflects the thermal activation of an additional fluctuating mechanism which becomes visible for O1 site, above $T^{\ast}$. Its prominent contribution, $(T_2^{-1})_a$, to $T_2^{-1}(T)$ is defined as the difference $(T_2^{-1})_a = T_2^{-1}(T) - T_2^{-1}(293~{\rm K})$. As seen in the inset of Fig.~\ref{Fig. 7.}, $(T_2^{-1})_a$ has an exponential behavior versus $1/T$ below 900~K. We model this mechanism with the correlation function $\langle h_{\perp}(0)h_{\perp}(t)\rangle=h_{\perp}(0)^2\exp(-t/\tau_c)$ which yields $(T_2^{-1})_a \propto h_{\perp}(0)^2\tau_c/[1+(\omega_L\tau_c)^2]$, where $\tau_c$ is the correlation time characterizing fluctuations of $h_{\perp}(t)$ and ${}^{17}\omega_L = 2\pi\nu_L$. Assuming that $\tau_c = \tau_{c0} \exp(E_a/T)$, the model reproduces the main experimental features i.e. for $\omega_L\tau_c\gg1$, $(T_2^{-1})_a$ scales $\tau_c^{-1}$ and $(T_2^{-1})_a$ reaches a maximum at $\omega_L\tau_c \sim 1$. The deduced energy barrier, $E_a = 0.76(8)$~eV, is consistent with estimates of $\Delta_{JT}$, the JT splitting of the $e_g$ state,  $0.7 < \Delta_{JT} <0.9~{\rm eV}$ \cite{Kovaleva_PRL93,Yin_PRL96,Pavarini_PRL104}.

As the deduced value, $\tau_{c0} =1.2(2)\cdot 10^{-12}$~sec, is too large to describe a vibrating $\rm MnO_6$ octahedron \cite{Granado_PRB62,Martin-Carron}, we speculate that some collective modes of adjacent octahedra \cite{Sanchez_PRL90,Qiu_PRL94} are responsible for the Mn spin slow fluctuations below and above $T_{JT}$. To date we are not able to specify all the parameters of these modes. Nevertheless, the propagation vector should be directed along the c-axis, providing above $T^{\ast}$ a melting mechanism of the AF spin correlations between adjacent Mn in this direction. Moreover, according to the $\rm {}^{139}La$ NMR results, the $\rm MnO_6$ octahedra should fluctuate in such a correlated manner that their averaged effect results in an asymmetric charge environment of La atoms (Fig.~\ref{Fig. 4.}). Even in the $O$-phase with a metrically cubic ($a \approx b \approx c/\sqrt{2}$) lattice \cite{Rodr-Carv_PRB57} the structural position of La remains aside the inversion symmetry point $\{0;0;0\}$ of the pseudo-cubic unit cell shown in Fig.~\ref{Fig. 1.}.

\section{Conclusion}

We have studied the long-range orbital order and its melting as well as the $\rm Mn^{3+}$--O--$\rm Mn^{3+}$ exchange interaction in the paramagnetic $O'$-phase  and above $T_{JT} =750$~K, in the $O$-phase of a stoichiometric $\rm LaMnO_3$ crystalline sample. 

At 293~K, the orbital mixing angle of the ground state wave function of the $e_g$-electron, $\varphi_{\rm nmr} =109\pm 1.5^{\circ}$, was obtained from $\rm {}^{17}O$ NMR. This value is close to $\varphi_{\rm str} =108^{\circ}$, deduced from structural data based on the Jahn-Teller distortions of the $\rm MnO_6$ octahedra \cite{Rodr-Carv_PRB57}. The fact that $\varphi_{\rm nmr}$ is close to $\varphi_{\rm str}$ supports the theoretical works \cite{Leonov_PRB81,Pavarini_PRL104,Yin_PRL96} which conclude that both, the SE interactions \textit{e--e} and the JT distortions \textit{e--l} are needed to explain the orbital ordering. In the orbital fluctuating $O$-phase the NMR data which correspond to a time averaged orbital configurations yield equally weighted orbitals for the $e_g$-doublet. Not so many experimental methods are able to yield the orbital mixing angle and this $\rm {}^{17}O$ NMR approach could be extended to other strongly correlated oxide materials with an active orbital degree of freedom.

The two distinct oxygen sites of the structure enable to probe different Mn--Mn spin correlations; O2 sites probe the Mn--Mn spin correlations in the \textit{ab}-plane whereas O1 sites probes the Mn--Mn spin correlations along the \textit{c}-axis. In the $O'$-phase, the NMR properties measured at O2 site are robust up to the transition temperature, $T_{JT}$, while at O1 site they show a marked change below $T_{JT}$, at about $T^{\ast}= 550$~K. Indeed, the ferromagnetic nature of the superexchange coupling, SE, between Mn spins in the \textit{ab}-plane is confirmed up to $T_{JT}$ whereas along the \textit{c}-axis, SE is antiferromagnetic up to $T^{\ast}$ and alters gradually toward FM-type with further heating, resulting in an anisotropic-to-isotropic change of the exchange coupling with reinforced ferromagnetic correlations above $T_{JT}$. Furthermore, the $\rm {}^{17}O$ spin-spin relaxation time, ${}^{17}T_2$, which shed light on the low frequency dynamics of the Mn spins, shows also different behavior at O2 and O1 sites. Indeed at O2 site, ${}^{17}T_2({\rm O2})$ is constant up to $T_{JT}$ while at O1 sites ${}^{17}T_2({\rm O1})$ data evidence that above $T^{\ast}$, slow fluctuations of the Mn spins along \textit{c} are thermally activated. Collective modes of adjacent octahedra along the \textit{c}-axis should be considered for this slow mechanism which yields changes of the spin correlations between adjacent Mn along \textit{c}, a spin marker of the orbital ordering. 

It is worth to note that at about 600~K, close to $T^{\ast}$, a change in slope of the thermal behavior of the resistivity of $\rm LaMnO_3$ was interpreted as the onset of orbital-disorder fluctuations \cite{Zhou_Goodenough_2003}. Nevertheless, static nanoscale charge and spin heterogeneities which might appear in such a scenario of the 3D-melting of the orbital order in $\rm LaMnO_3$ when approaching $T_{JT}$ from below are not confirmed by $\rm {}^{139}La$ NMR. Our $\rm {}^{17}O$ NMR results show that $T^{\ast}$ marks rather a 3D to 2D crossover since only the magnetic coupling between Mn neighbors in adjacent \textit{ab}-layers alters and thus $T^{\ast}$ marks the onset of a melting of the long-range orbital order along the \textit{c}-axis.

\medskip
\smallskip
\smallskip
\smallskip
\smallskip

\textbf{ACKNOWLEDGEMENTS}
\medskip
\smallskip
\smallskip

We acknowledge Dm. Korotin for fruitful discussion of the DMFT results in $\rm LaMnO_3$. This work is supported in part by the RFBR Grant No~12-02-00358 and the UB RAS Research Projects
Ns 12-M-23-2061; 12-Y-2-1025. S.V. and A.G. thank ESPCI for hospitality and support.


\end{document}